\documentstyle[twoside,fleqn,espcrc2,epsfig]{article}

\newcommand{\AmS}{{\protect\the\textfont2
  A\kern-.1667em\lower.5ex\hbox{M}\kern-.125emS}}

\title{Mott transition in one dimension}

\author{T. Giamarchi
\address{Laboratoire de Physique
Des Solides, U.P.S. B\^at 510, 91405 Orsay, France}
\thanks{Laboratoire associ{\'e} au CNRS. Email: giam@lps.u-psud.fr}
}

\begin{document}

\begin{abstract}
I review some of the results on the Mott transition in one dimensional
systems obtained in
\protect{\cite{giamarchi_umklapp_1d,giamarchi_attract_1d,giamarchi_curvature}}.
I discuss the phase diagram and critical properties of
both Mott transitions at fixed filling and upon doping, as well as the
dc and ac conductivity. Application of these results to organic
conductors is discussed.
\end{abstract}

\maketitle

\section{Introduction}

The Mott transition is one of the most fascinating phenomenons arising
from electron-electron interactions, and occurs in a wide range of
materials \cite{mott_metal_insulator}.
In fact two different Mott transitions exist: one can
either stay at a given (commensurate) filling and vary the strength of
the interactions (I call this transition Mott-U and it occurs in e.g.
vanadium oxides or
in organic (quasi-)one dimensional systems),  one can also keep the
strength of the interaction constant and dope the system to move away
from the commensurate density (a Mott-$\delta$ transition, a situation
realized in High Tc superconductors or in quantum wires).
Although the basic underlying physics behind these transitions is by
now well understood it has proved incredibly difficult to tackle it in
either $d=2$ or $d=3$ due to our lack of tools to treat strongly
interacting systems \cite{mott_mean_fields}. In fact nearly all the fine
points of the transition, such as the critical properties or the
transport properties remain unknown.

One dimension constitute a special case where a rather complete study of
the Mott transition can be done.  This offers special
interest both for theoretical and experimental reasons. From a
theoretical point of view, the effect of e-e interactions is
particularly strong and leads to a non-fermi
liquid state (the so called Luttinger liquid (LL)).
One can therefore expect drastic effects on the transport properties of
the system. From the experimental point of view, both transitions at
constant doping and by varying the doping can be realized, e.g. in
organic conductors \cite{jerome_revue_1d} and quantum wires
\cite{tarucha_quant_cond} or Josephson junction networks
\cite{vanoudenaarden_josephson_mott}.

Although the thermodynamic properties of the Mott transition
were understood a long time ago for the Hubbard model, which was shown
to be a Mott insulator at half
filling \cite{lieb_hubbard_exact,emery_revue_1d,solyom_revue_1d},
very little  was
known of the Mott-$\delta$ transition and of the transport
properties: a parquet treatment gave the effective scattering
\cite{gorkov_pinning_parquet} but was limited to half filling and small
(perturbative) interactions, and only the zero frequency
conductivity (i.e.
the Drude weight) could be computed by Bethe-Ansatz for the particular
case of the Hubbard model
\cite{shastry_twist_1d,schulz_conductivite_1d}).
Recently a complete picture of both
Mott transitions as well as a full description of the transport
properties $\sigma(\omega,T,\delta)$ for
any commensurate filling, both for bosons or fermions,
was obtained
\cite{giamarchi_umklapp_1d,giamarchi_attract_1d,giamarchi_curvature}.
In these proceedings I will review some of these results.
No derivation will be given and
the reader is referred to
\cite{giamarchi_umklapp_1d,giamarchi_attract_1d,giamarchi_curvature}
for derivation as well as the complete
analytical expressions, for the figures presented here. Such a
presentation is done in section~\ref{section2} where umklapp effects are
presented and in section~\ref{section3} where both the critical
properties of the Mott transition(s) and the transport properties are
examined. In section~\ref{section4} application of these results to the
physics of the organic materials is done.

\section{Lattice effects and umklapp terms} \label{section2}

In the continuum  e-e interactions conserve momentum, and thus
current,  and  cannot lead to any finite conductivity as
a consequence of Galilean invariance. In the presence of
a lattice however the momentum need only to be conserved modulo one
vector of the reciprocal lattice, and such interaction process (named
umklapp process) can lead to finite resistivity. In a fermi liquid,
umklapps are responsible for the intrinsic resistivity $\rho(T)\sim T^2$.

In one dimension it was rapidly
realized \cite{emery_revue_1d,solyom_revue_1d}
that umklapps are also responsible
for the Mott-U transition at half filling. Away from half filling they
are ``frozen'' due to the mismatch in momentum, and are usually
discarded as irrelevant: the system becomes then a perfect metal.
However both for the transport properties and
to study the Mott-$\delta$ transition it is necessary to
have a description of the umklapp processes even for finite doping.

This can be achieved using the so called bosonization representation,
that uses that {\bf all} excitations of a one dimensional
system can be described in term of density oscillations
\cite{emery_revue_1d,solyom_revue_1d,haldane_bosonisation}.
The charge properties of a
{\bf full} interacting one dimensional system (excluding umklapp terms)
is therefore described by
\begin{equation} \label{quadra}
H_0 = \frac1{2\pi} \int dx \; u_\rho K_\rho (\pi\Pi_\rho)^2 +
                 \frac{u_\rho}{K_\rho} (\nabla \phi)^2
\end{equation}
where $\nabla \phi = \rho(x)$, $\rho(x)$ is the charge density and $\Pi$
is the conjugate momentum to $\phi$. All the interaction effects are
hidden in the parameters $u_\rho$ (the velocity of charge excitations)
and $K_\rho$ (the Luttinger liquid exponent controlling the decay of
all correlation functions). This description (\ref{quadra}) is
valid for an arbitrary
one-dimensional interacting system, {\bf provided} one uses the proper
$u$ and $K$ (in the following I will drop the $\rho$ index).
In a general way $K =1$ is the
noninteracting point, $K>1$ means attraction whereas $K<1$ means
repulsion.

The umklapp process can also be given in terms of boson operators
\cite{emery_revue_1d,solyom_revue_1d}. In fact umklapps exist not only
at half filling but for higher
commensurabilities as well by transferring more particles across the fermi
surface (such processes are generated in higher order in perturbation
theory) \cite{giamarchi_curvature,schulz_mott_revue}.
For even commensurabilities the
Hamiltonian corresponding to the umklapp process is
\begin{equation} \label{um1n}
H_{\frac1{2n}} = g_{\frac1{2n}} \int dx \cos(n \sqrt8 \phi_\rho(x) +
\delta x)
\end{equation}
where $n$ is the order of the commensurability ($n=1$ for half filling -
one particle per site; $n=2$ for quarter filling - one particle every
two sites and so on). The coupling constant $g_{1/2n}$ is the umklapp
process corresponding to the commensurability $n$ and $\delta$ the
deviation (doping) from the commensurate filling.
$g_{\frac12}$ corresponds to one particle per site (half filling)
For simple models such as the Hubbard model $g_{\frac12}=U$, but
this does not need to be the case for more general models (see in
particular section~\ref{section4}). For $1/4$ filling and a typical
interaction $U$ one has $g_{1/4}\sim U (U/E_F)^3$.
Odd commensurability involves spin \cite{schulz_mott_revue} but can be
treated similarly.
Similar expressions can be derived for the case of bosons
\cite{giamarchi_attract_1d}.
It is therefore remarkable that in one dimension $H_0+H_{\frac1{2n}}$
provides the solution to {\bf all} Mott transitions,
for {\bf all} systems and {\bf all} (for particles with spin: even) 
commensurabilities.

\section{Mott transition(s)} \label{section3}

Let us now examine the physical properties of the Mott transitions close
to a commensurability of order $n$ described by $H_0+H_{\frac1{2n}}$.

\subsection{Phase diagram}

\begin{figure}
   \centerline{\epsfig{file=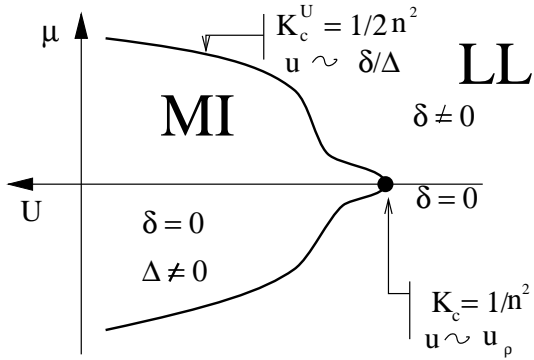,angle=-90,width=7cm}}
\caption{Phase diagram close to a commensurability of order $n$ ($n=1$
for half filling and $n=2$ for quarter filling). $U$ denotes a general
(i.e. not necessarily local) repulsion. $\mu$ is the chemical potential
and $\delta$ the doping. MI means Mott insulator and LL Luttinger
liquid (metallic) phase. The critical exponent $K_c$ and velocity $u$
depends on whether it is a Mott-U or Mott-$\delta$ transition.}
\label{phasediag}
\end{figure}

The Mott-U and Mott-$\delta$ transitions are radically different and
lead to the phase diagram shown in Figure~\ref{phasediag}.

The Mott-U is of the Kosterlitz Thouless type
\cite{emery_revue_1d,solyom_revue_1d} and
occurs for a critical value of $K$, $K_c=1/n^2$
\cite{giamarchi_curvature,schulz_mott_revue}. For
half filling the transition point is the noninteracting one ($K_c=1$)
but for higher commensurability one reaches the Mott insulator only for
very repulsive interactions (for example for quarter filling
$K_c^U=1/4$). In the metallic phase the system is a LL, with finite
compressibility and Drude weight. The Mott insulator has a gap in the
charge excitations (thus zero compressibility). At the transition there
is a finite jump both in the compressibility and Drude weight.
The dynamical exponent is $z=1$.

\begin{figure}
   \centerline{\epsfig{file=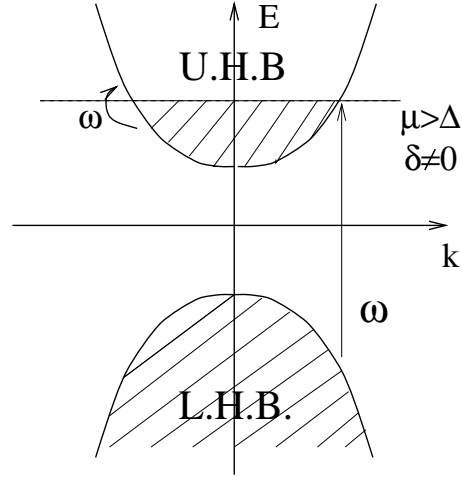,angle=-90,width=6cm}}
\caption{Lower Hubbard band and Upper Hubbard band. This concept can be
made rigorous in one dimension by mapping the full interacting system to
a massive Thiring model. Optical transitions can be made either within
or between the two ``bands''.}
\label{Thiring}
\end{figure}

To study the Mott-$\delta$ transition it is useful
\cite{giamarchi_umklapp_1d} to map the
sine-gordon Hamiltonian $H_0+H_{\frac1{2n}}$ to a spinless fermion model
(known as massive Thiring model \cite{emery_revue_1d,solyom_revue_1d}),
describing the charge excitations
(solitons) of the sine-gordon model. The remarkable fact is that {\bf
close} to the Mott-$\delta$ transition the solitons become
non-interacting, and one is simply led to a simple semi-conductor
picture of two bands separated by a gap (see figure~\ref{Thiring}).
This image has to be used with caution since the solitons are only
non-interacting for infinitesimal doping (or for a very special value
of the initial interaction) and has to be supplemented by other
techniques \cite{giamarchi_umklapp_1d}. Nevertheless it provides a very
appealing description of the LHB and UHB and a good guide to understand
the phase diagram and transport properties.
The Mott-$\delta$ transition is of
the commensurate-incommensurate type. The {\bf universal}
(independent of the interactions) value of the
exponents $K_c^\delta = 1/(2n^2)$ is half of the one of Mott-U
transition. Since at the Mott-$\delta$ transition the
chemical potential is at the bottom of a band the velocity goes to zero
with doping. This leads to a continuous vanishing of the Drude weight
and compressibility. The dynamical
exponent is now $z=2$. For more details see
\cite{giamarchi_umklapp_1d,giamarchi_curvature,schulz_mott_revue,mori_mott_1d}.
For bosons, the phase diagram of
figure~\ref{phasediag} is well compatible with numerical results
\cite{batrouni_bosons_numerique} and higher dimensional proposals
\cite{fisher_boson_loc}.  

\subsection{Transport properties}
Let us now look at the transport properties. The full conductivity
(real and imaginary part) $\sigma(\omega,T,\delta)$ can be found in
\cite{giamarchi_umklapp_1d,giamarchi_attract_1d,giamarchi_curvature} and
we just examine here simple limits.

\begin{figure}
   \centerline{\epsfig{file=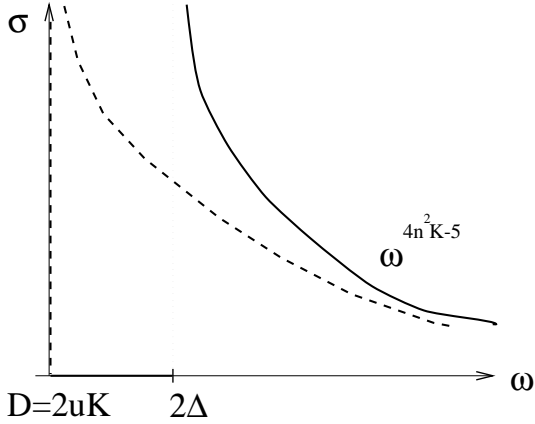,angle=-90,width=7cm}}
\caption{ac conductivity for $\delta=0$ for a commensurability of order
$n$. $\Delta$ is the Mott gap. The full line is the conductivity in the
Mott insulator. The dashed one is $\sigma$ in the metallic
regime. It contains both a Drude peak of weight $D$ and a regular part.}
\label{sigacdel0}
\end{figure}

The ac conductivity (at $T=0$) for
$\delta=0$ is shown in figure~\ref{sigacdel0}.
In the Mott insulator
$\sigma$ is zero until $\omega$ can make transitions between the LHB and
UHB. At the threshold one has the standard square root singularity
coming from the density of states (see figure~\ref{Thiring}). For higher
frequencies interactions dress the umklapps and give a
nonuniversal (i.e. interaction-dependent) power law-like decay. Such
a power law is beyond the reach of the simple noninteracting description
of Figure~\ref{Thiring}.

\begin{figure}
   \centerline{\epsfig{file=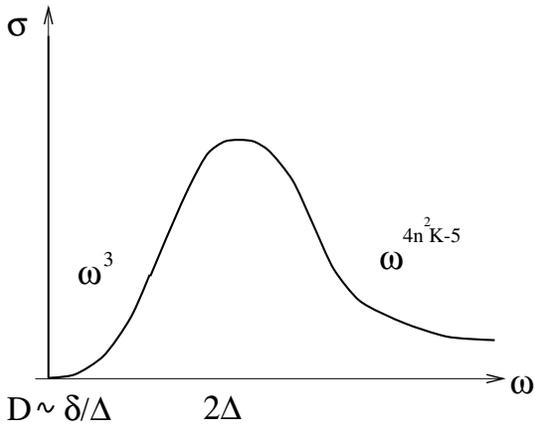,angle=-90,width=7cm}}
\caption{ac conductivity for $\delta\ne 0$ for a commensurability of
order $n$. $\Delta$ is the Mott gap. In addition to the Drude peak of
weight $D$ the regular part has two distinct regimes. }
\label{sigacdeln0}
\end{figure}

Away from commensurate filling ($\delta\ne 0$) the conductivity is shown
in Figure~\ref{sigacdeln0} (only the case where the half filled system
is a MI is shown. For the other case see \cite{giamarchi_attract_1d}).
Features above the Mott gap are unchanged (the system has no way to
know it is or not at half filling at high frequencies). The two new
features are a Drude peak with a weight proportional to $\delta/\Delta$,
and an $\omega^3$ absorption \cite{giamarchi_curvature} at small
frequency. Features above the Mott
gap come from inter (hubbard)-band transitions whereas they come from
intra-band processes below the Mott gap (see figure~\ref{Thiring}).

\begin{figure}
   \centerline{\epsfig{file=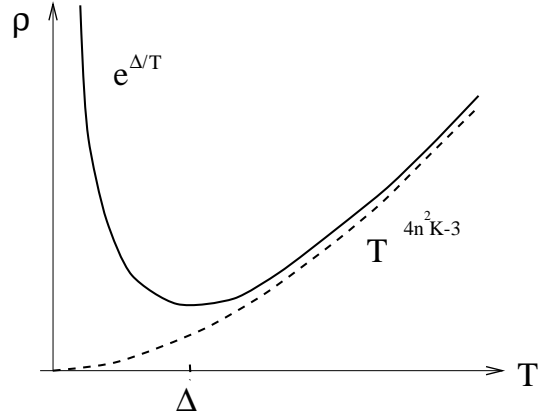,angle=-90,width=7cm}}
\caption{dc conductivity as a function of $T$. Full line is for the
Mott insulator, dashed line is in the metallic regime. $\Delta$ is the
Mott gap.}
\label{dcsig}
\end{figure}

The dc conductivity can be computed by the same methods and is shown in
figure~\ref{dcsig}.
Here again the dressing of umklapps by the other interactions
results in a nonuniversal power law dependence. If the interactions are
repulsive enough the resistivity can even {\bf increase} as a function
of temperature well above the Mott gap.

Two universal behavior are expected: at the Mott-U transition one has
$\rho(T)\sim T$ and $\sigma(\omega)\sim 1/(\omega\ln(\omega)^2)$,
whereas at the Mott-$\delta$ transition due to the different $K_c$ one
expects $\rho(T)\sim 1/T$.

All this results are completely general and apply to any
one-dimensional systems for which $\Delta$ is smaller than the scale
above which {\bf all} interactions can be treated perturbatively
(typically $U$), a situation that covers most of
the experimentally relevant cases (see section~\ref{section4}).
It is noteworthy that the above results are also valid in the
presence not of umklapp processes but of a simple periodic potential
(the lattice corresponds itself to a $4k_F$ periodic potential). For a
$2k_F$ periodic potential transport properties are similar to the one
above with the replacement of $4n^2 K_\rho$ by $1+K_\rho$.

\section{Organic compounds} \label{section4}

The above results have a direct application to
organic conductors. These compounds are $1/4$ filled by chemistry but
due to a slight dimerization of the chain an half
filled umklapp $g_{1/2}\sim U (D/E_F)$ also exists where $D$ is the
dimerization gap, and $U$ a typical strength of the interactions
\cite{jerome_revue_1d}. Since $D/E_F$ is quite small the
umklapp term is much smaller than the other interactions leading to a
quite small Mott gap (see e.g. \cite{penc_numerics} for a numerical
estimation of the parameters). There is also a
$1/4$ filled umklapp
$g_{1/4}\sim U (U/E_F)^3$, which is as we saw less relevant but can be
depending on the typical interaction $U$ much larger in magnitude than
$g_{1/2}$.

\begin{figure}
   \centerline{\epsfig{file=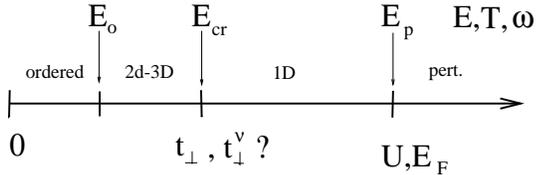,angle=-90,width=7cm}}
\caption{Four important energy regimes for quasi-one dimensional
systems. In ``pert.'' everything can be treated perturbatively. In
``1D'' the
interactions lead to the one dimensional physics, and hopping from chain
to chain is incoherent. In ``2D-3D'' the hoping between chains is
coherent. The system orders in ``Ordered''.}
\label{scales}
\end{figure}

Since the organic conductors are only quasi-one dimensional systems with
a perpendicular hopping integral $t_\perp$ between the chains one can
distinguish various domains in energy scale (temperature or
frequency) as shown in figure~\ref{scales}
The most relevant questions being of course: what is the strength of the
interactions in these systems, what is the scale for $T_{\rm cr}$ (the
bare $t_\perp$ or lower \cite{hopping_general}), and what is the physics
below $T_{\rm cr}$.

In the absence of $t_\perp$ one expects therefore these compounds to be
Mott insulators. This is the case for the TMTTF family that has indeed
a conductivity \cite{creuzet_tmttf} similar to the one of
figure~\ref{dcsig} (full line).
Indeed if $\Delta > T_{\rm cr}$, one expects the Mott gap to render the
single particle hopping $t_\perp$ irrelevant ($E_{\rm cr}$ would thus
not exist). This family should
be described by one-dimensional physics. Further check of this can be
provided by examination of the optical (ac) conductivity, and comparing
it to figure~\ref{sigacdel0}. Measurements of the transverse
conductivity would also give information on the relevance of the
transverse hopping. Note that the temperature dependence of
the dc resistivity and the frequency dependence of the optical
conductivity provide a {\bf direct} measure of the $K_\rho$ exponent of
the Luttinger Liquid and give therefore crucial information on the
importance of interactions in such systems (the optical conductivity
has the advantage to be free from thermal expansion problems).
A naive fit in TMTTF would give a value of $K_\rho\sim 0.8$, widely
different from the one of $K_\rho=0.3$ extracted from the NMR
\cite{wzietek_tmtsf_nmr}. A way
to get out of this predicament could be that the conductivity is in fact
dominated by $1/4$ filling umklapp processes till very close to the Mott
gap giving
\begin{equation}
\rho(T) \sim g_{1/2}^2 T^{4K-3} + g_{1/4}^2 T^{16K-3} \sim g_{1/4}^2 T^{16K-3}
\end{equation}
but this point clearly deserves further investigation.

On the other hand, the TMTSF family shows
a rather good metallic behavior with a $T^2$ resistivity, indicating the
importance of transverse hopping. This is to be expected if
$\Delta > E_{\rm cr}$. There is important controversy on the value of
$E_{\rm cr}$ \cite{behnia_transport_magnetic,gorkov_sdw_tmtsf}.
Regardless of the value of $E_{\rm cr}$ the physics for
$(\omega, T ) > E_{\rm cr}$  will still
be controlled by one dimensional
effects. Indeed For the TMTSF family the optical conductivity
\cite{dressel_optical_tmtsf} is
very well compatible with the figure~\ref{sigacdeln0}.
In particular the optical peak can easily be interpreted in term of the
Mott insulator described here. Of course more detailed comparison of the
structure above the gap an in particular a check for the power law decay
of figure~\ref{sigacdeln0} would be worthy to do. The low energy
features and in particular the metallic behavior are closer to the
{\bf doped} system rather than the commensurate one.
``doping'' is not
too surprising since if one particle hopping between chains is relevant, one
expect small deviations to the commensurate filling due to the warping
of the Fermi surface. One therefore expect a very small spectral weight
in the $\delta(\omega)$ part.
Since one has a clear idea of the (purely)
one-dimensional conductivity (figure~\ref{sigacdeln0}), a detailed
comparison with experimental data should provide an indication on the
value of $E_{\rm cr}$. The question on whether the physics below
$E_{\rm cr}$ is simply ``fermi liquid'' like
\cite{gorkov_sdw_tmtsf} or still retains some
features of one-dimensionality and interactions is still open. 
One way to settle this issue is a detailed examination of the 
low frequency part of the optical conductivity and measurements of the
transverse dc conductivity in this regime.
Another way would
be to examine the effects of impurities on the dc conductivity. Indeed
one expect drastic localization in a one dimensional regime and very
weak effects for a FL \cite{giamarchi_loc}.

{\bf Acknowledgments:}

It is a pleasure to thank L. Degiorgi, L.P. Gor'kov, G. Gr\"uner, 
D. J\'erome, A.J. Millis, H.J. Schulz and B.S. Shastry   
for many interesting discussions.


\begin{thebibliography}{10}

\bibitem{giamarchi_umklapp_1d}
T. Giamarchi, Phys. Rev. B {\bf 44},  2905  (1991).

\bibitem{giamarchi_attract_1d}
T. Giamarchi, Phys. Rev. B {\bf 46},  342  (1992).

\bibitem{giamarchi_curvature}
T. Giamarchi and A.~J. Millis, Phys. Rev. B {\bf 46},  9325  (1992).

\bibitem{mott_metal_insulator}
N.~F. Mott, {\em Metal-Insulator Transitions} (Taylor and Francis, London,
  1990).

\bibitem{mott_mean_fields}
Various mean field theories like slave bosons or $d=\infty$ limit can be
  applied to study the Mott transition in more than one dimension. See e.g. A.
  Georges, G. Kotliar, W. Krauth and M. J. Rozenberg Rev. Mod. Phys. {\bf
  68} 13 (1996) and references therein.

\bibitem{jerome_revue_1d}
D. J\'erome and H.~J. Schulz, Adv. Phys. {\bf 31},  299  (1982).

\bibitem{tarucha_quant_cond}
S. Tarucha, T. Honda, and T. Saku, Sol. State Comm. {\bf 94},  413  (1995).
S. Tarucha, T. Saku, Y. Tokura, and Y. Hirayama, Phys. Rev. B {\bf 47},  4064
  (1993).

\bibitem{vanoudenaarden_josephson_mott}
A. {van Oudenaarden} and J.~E. Mooij, Phys. Rev. Lett. {\bf 76},  4947  (1996).

\bibitem{lieb_hubbard_exact}
E.~H. Lieb and F.~Y. Wu, Phys. Rev. Lett. {\bf 20},  1445  (1968).

\bibitem{emery_revue_1d}
V.~J. Emery,  in {\em Highly Conducting One-Dimensional Solids}, edited by
  J.~T.~D. et~al. (Plenum, New York, 1979), p.\ 327.

\bibitem{solyom_revue_1d}
J. S\'olyom, Adv. Phys. {\bf 28},  209  (1979).

\bibitem{gorkov_pinning_parquet}
L.~P. Gorkov and I.~E. Dzyaloshinski, JETP Lett. {\bf 18},  401  (1973).

\bibitem{shastry_twist_1d}
B.~S. Shastry and B. Sutherland, Phys. Rev. Lett. {\bf 65},  243  (1990).

\bibitem{schulz_conductivite_1d}
H.~J. Schulz, Phys. Rev. Lett. {\bf 64},  2831  (1990).

\bibitem{haldane_bosonisation}
F.~D.~M. Haldane, J. Phys. C {\bf 14},  2585  (1981).

\bibitem{schulz_mott_revue}
H.~J. Schulz,  in {\em ``Strongly correlated electronic materials''}, 
edited by K.~S. Bedell and al. (Addison-Westley,
Reading, Massachusetts, 1994).

\bibitem{mori_mott_1d}
M. Mori, H. Fukuyama, and M. Imada, J. Phys. Soc. Jpn. {\bf 63},  1639  (1994).

\bibitem{batrouni_bosons_numerique}
G.~G. Batrouni, R.~T. Scalettar, and G.~T. Zimanyi, Phys. Rev. Lett. {\bf 65},
  1765  (1990).

\bibitem{fisher_boson_loc}
M.~P.~A. Fisher, P.~B. Weichman, G. Grinstein, and D.~S. Fisher, Phys. Rev. B
  {\bf 40},  546  (1989).

\bibitem{penc_numerics}
K. Penc and F. Mila, Phys. Rev. B {\bf 50},  11429  (1996).

\bibitem{hopping_general}
See e.g. D. Boies, C. Bourbonnais and A.-M. S. Tremblay Phys. Rev. Lett {\bf
  74} 968 (1995) and references therein.

\bibitem{creuzet_tmttf}
F. Creuzet et al., J. Phys. (Paris) {\bf C3},  1099  (1983).

\bibitem{wzietek_tmtsf_nmr}
P. Wzietek et al., J. Phys. (Paris) {\bf 3},  171  (1993).

\bibitem{behnia_transport_magnetic}
K. Behnia et al., Phys. Rev. Lett. {\bf 74},  5272  (1995).

\bibitem{gorkov_sdw_tmtsf}
L. P. Gor'kov ``Non fermi liquid features in Q1D organic conductors due to
  their proximity to SDW state.'' this volume.

\bibitem{dressel_optical_tmtsf}
M. Dressel et al., Phys. Rev. Lett. {\bf 77},  398  (1996).

\bibitem{giamarchi_loc}
T. Giamarchi and H.~J. Schulz, Phys. Rev. B {\bf 37},  325  (1988) and
references therein.

\end{thebibliography}

\end{document}